\documentclass[aps,pra,twocolumn,superscriptaddress, biblatex]{revtex4-2}

\pdfoutput=1
\usepackage[english]{babel}

\usepackage[colorlinks=true, breaklinks=true]{hyperref}
\usepackage{epsfig,graphicx}
\usepackage{indentfirst}
\usepackage{amsfonts,amssymb,latexsym,amsmath,enumerate,amsthm}
\usepackage{bm}
\usepackage{color}
\usepackage{appendix}

\usepackage{dcolumn}
\usepackage{centernot}

\usepackage{courier}

\usepackage{url}

\usepackage{physics}
\usepackage{mathtools}
\DeclareMathOperator{\T}{T}
\usepackage{soul}

\usepackage[sort&compress]{natbib}
\begin{document}

\title{Vacuum polarization and Wichmann-Kroll correction in the finite basis set approximation}

\author{V.\,K.~Ivanov}
    \email[E-mail: ]{vladislav.ivanov@metalab.ifmo.ru}
\affiliation{School of Physics and Engineering, ITMO University, 
197101 St.\,Petersburg, Russia}

\author{ S.\,S.~Baturin}
\affiliation{School of Physics and Engineering, ITMO University, 
197101 St.\,Petersburg, Russia}

\author{ D.\,A.~Glazov}
\affiliation{School of Physics and Engineering, ITMO University, 
197101 St.\,Petersburg, Russia}

\author{ A.\,V.~Volotka}
\affiliation{School of Physics and Engineering, ITMO University, 
197101 St.\,Petersburg, Russia}

\date{\today}

\begin{abstract}
The finite basis set method is commonly used to calculate atomic spectra, including QED contributions such as bound-electron self-energy. Still, it remains problematic and underexplored for vacuum-polarization calculations. We fill this gap by trying this approach in its application to the calculation of the vacuum-polarization charge density and the Wichmann-Kroll correction to the electron binding energy in a hydrogen-like ion. We study the convergence of the method with different types and sizes of basis sets. We cross-check our results for the Wichmann-Kroll correction by direct integration of the Green's function. As a relevant example, we consider several heavy hydrogen-like ions and evaluate the vacuum polarization correction for $S$ and $P$ electron orbitals.

\end{abstract}

\maketitle

\section{Introduction}

Quantum electrodynamics (QED) is one of the most established and well-developed field theories. Extensive experimental tests show a high predictive power of this theory and an almost extreme accuracy. The latter is due to smallness of the fine-structure constant $\alpha\approx1/137$, which is the coupling constant of the theory. To date, many experimental tests of QED have been performed. However, there is still considerable interest and development when it comes to QED in strong electromagnetic fields. Such processes reveal tiny effects that are not observed in the regular scenario and allow testing QED with much higher precision. 
In highly charged heavy ions 
, it is remarkable that the electric field is only a few orders smaller than the Schwinger limit. For example, the surface field of the Uranium nucleus is $\sim 10^{19} \text{ V/cm}$. This makes highly charged heavy ion a perfect system for testing the strong field QED.  

To perform these tests and to compare theory with experiment, a precise calculation of the lowest electron level binding energies is required. A natural approach to the analysis of different systems within QED is the perturbation series, which could be evaluated with some effort to a high order in small parameter $\alpha$ and thus to a high degree of accuracy. While algorithms exist to perform such calculations, many of them require significant computational resources, especially when it comes to the evaluation of the higher order terms. Optimization and improvement of the computational methods and algorithms as well as increasing the computational speed is one of the prominent research directions.

Development of such algorithms becomes especially important to calculate QED effects in systems in the strong field regime. In such systems the high-precision QED calculations become complex since the parameter, by which an electron is coupled to the external field, may be large. For instance, in the presence of a heavy nucleus, the electrons are coupled to the external field through the factor $Z\alpha$, where $Z$ is the atomic number -- this coupling constant is not sufficiently small when considering heavy elements with $Z\sim100$. For such systems, the contribution of higher-order terms in the $Z\alpha$ expansion becomes too large, thus perturbative methods, mentioned above, cannot be used. \cite{Mohr1998}. To address this issue, it is necessary to employ the non-perturbative approach, known as the Furry's picture, in which the external field of the nucleus is incorporated into the electron wave function.

\begin{figure}[t]
    \centering
    \includegraphics[width=1\linewidth]{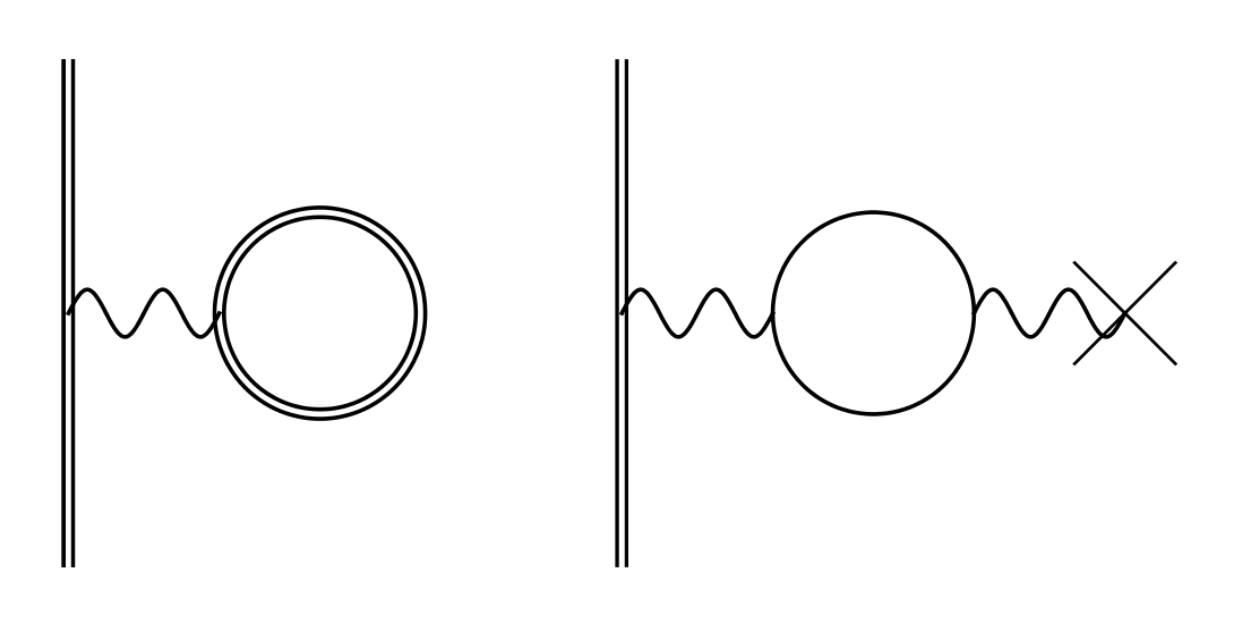}
    \caption{Vacuum polarization diagrams. Left: vacuum polarization in Furry's picture, all orders of $\alpha(Z\alpha)$ are included. Right: the Uehling correction. Double lines denote an electron in an external field, single straight lines are free electron propagators, cross is the external field source, wavy lines denote a propagating photon.
    }
    \label{fig:vp_diag}
\end{figure}

In this work, we focus on the vacuum polarization effect (Fig. \ref{fig:vp_diag}), which along with the one-photon exchange (relevant for atoms with more than one electron) and the self-energy diagrams consist the main QED contributions to atomic spectrum. Vacuum polarization (VP) can be thought of as the creation of a virtual electron-positron pair under the influence of the strong nuclei potential on the Dirac sea. This interaction effectively attracts the electron part of the electromagnetic vacuum while pushing away the positron part. As a consequence of the aforementioned local charge imbalance, the nucleus field is effectively screened. VP leads to a Lamb shift in the atomic spectrum \cite{Mohr1998, Yerokhin2019}
and, together with other types of QED contributions, constitutes an indispensable component of the g-factor calculation \cite{Karshenboim2001, Yerokhin_2013, belov2016muonic, Dizer2023}.  The calculations of the g-factor of highly charged ions are currently being actively explored \cite{Volotka2013}, as they have a number of important applications. These include the precise determination of the electron mass by studying hydrogen-like ions \cite{Sturm2014}, as well as a possible independent test for the value of the fine-structure constant \cite{Shabaev2006, Yerokhin2016}. In addition to hydrogen-like ions, systems with a larger number of electrons are being considered. These include highly charged two-electron ions \cite{Artemyev1997} and beryllium-like ions \cite{Malyshev2014}. Furthermore, vacuum polarization with more than one loop \cite{Yerokhin2019} is of interest.

In the leading order in $Z\alpha$, the VP contribution induces the Uehling potential. It provides the major part of the energy shift in the atomic spectrum originating from vacuum polarization (see, for example, \cite{Persson1993}). 
For heavy elements, higher orders in $Z\alpha$ expansion become significant. 
The $\alpha(Z\alpha)^3$ contribution was calculated by Wichmann and Kroll \cite{Wichmann1956}, and later in \cite{Gyulassy1974}, many-potential contribution from $\alpha(Z\alpha)^{n \geq 3}$ was also found. 
The importance of the Wichmann-Kroll vacuum polarization effect become apparent for heavy hydrogen-like ions, for which the VP Lamb shift is on the order of eV \cite{Mohr1998, Persson1993}.

In the present study, we evaluate the VP charge density using the finite basis set (FBS) approximation. This approach has already been used for electron self-energy diagrams \cite{Shabaev_DKB}. We take the next step and explore its applicability and accuracy in connection to vacuum polarization. The standard approach to these diagrams is to integrate the Green's function directly, which is a time-consuming and complicated procedure. The first step towards developing this procedure was discussed in \cite{Salman2023}, where the FBS method was applied to evaluate the vacuum polarization density on a Gaussian basis set. We build on these findings by testing the FBS method with both the Gaussian basis set and the B-spline basis set. The latter is widely used for self-energy diagrams. We calculate the induced charge density for various partial wave contributions up to the value of $\abs{\kappa} = 5$ and obtain the corresponding Wichmann-Kroll corrections for the electron energy levels of several hydrogen-like ions.  

Throughout this paper, we use a natural unit system $\hbar = c = m = 1$.

\section{Basic theory}\label{section_ii}

\subsection{Radial Dirac equation}

We start from a single electron Dirac equation that has the form:

\begin{equation}
    (\gamma^\mu p_\mu - m) \psi = 0,
\end{equation}

\noindent where $\gamma^\mu$ are the Dirac gamma matrices, 
$p_\mu$ is the four-momentum of the electron, and $m$ is the electron mass. Further, we assume $m=1$. For a free electron, $p_\mu$ 
is the kinetic momentum $p_\text{kin}$, and for an electron in an external field, 
$(p_\text{kin} - eA)$, 
where $e$ is the electron charge and $A$ is the external photon field potential. 

For the fermionic field operator, we can write 
\begin{equation}
    \psi(x) = \sum\limits_{E_n > 0 } a_n \phi_n(x) + \sum\limits_{E_n < 0 } b^\dagger_n \phi_n(x),
\end{equation}

\noindent where $a_n, b_n^\dagger$ are the annihilation/creation operators for an electron/positron and $\phi_n(x)$ are fermionic wave functions. We assume the spectrum 
to be discrete, which 
is customary in bound-state QED calculations.

For time-independent problem of 
an electron in an external electric field, the Dirac equation 
for the wave functions $\phi_n(x)$ can be written in the following form:

\begin{equation}
    h_D \phi_n (x) = E_n \phi_n(x),
\label{dirac_eq_2}
\end{equation}

\noindent where the Hamiltonian $h_D$ is expressed as

\begin{align}
    h_D =& -i\vb{\alpha}\nabla + \beta + V(\vb{x}), \label{wave_1}\\    
    \beta =& \gamma^0,\, \vb{\alpha} = \gamma^0 \vb{\gamma}.
\label{wave_2}
\end{align}

\noindent Here, $\nabla$ is a standard nabla operator, and $V(x)$ is the external field potential (the potential of the nucleus in our case).


Stationary solutions to the Dirac equation can be expressed as

\begin{equation}
    \phi_n(x) = \phi_n(\vb{x}) e^{-iE_n t}.
\end{equation}

For a spherically symmetric problem, we can perform a separation of variables and
achieve a radial Dirac equation for the radial component of wave functions (\ref{dirac_eq_2}), which also provides the electron energy spectrum. 
To separate the variables, we choose the wave function in the form

\begin{equation}
    \phi_n (\vb{x}) = \frac{1}{r}
    \begin{bmatrix}
        P_{n,\kappa}(r) \Omega _{\kappa,m}(\theta,\varphi) \\
        i Q_{n,\kappa}(r) \Omega _{-\kappa,m}(\theta,\varphi)
    \end{bmatrix},
\label{anzatz}
\end{equation}

\noindent where $\kappa$ is the relativistic angular quantum number, 
related to the total and orbital momenta as $j=\abs{\kappa} - \frac{1}{2}$ and $l=\abs{\kappa + \frac{1}{2}} - \frac{1}{2}$, respectively. 
The description of spherical spinors $\Omega(\theta,\varphi)$ can be found, for example, in \cite{Mohr1998, berestetskii, grant2007relativistic}). We substitute the ansatz (\ref{anzatz}) into the Dirac equation (\ref{dirac_eq_2}), and arrive at the radial Dirac equation:

\begin{equation}
    \begin{bmatrix}
        m + V & -\frac{d}{dr} + \frac{\kappa}{r} \\
        \frac{d}{dr} + \frac{\kappa}{r} & - m + V
    \end{bmatrix}
    \begin{bmatrix}
        P_{n,\kappa} \\
        Q_{n,\kappa}
    \end{bmatrix}
    =
    E_n \begin{bmatrix}
        P_{n,\kappa} \\
        Q_{n,\kappa}
    \end{bmatrix}.
    \label{dirac_eq_mat}
\end{equation}

\noindent The equation above has an analytic solution in the case of the Coulomb potential \cite{berestetskii}. For an arbitrary potential, one can still proceed numerically. Variational methods are the most common choice for numerical solutions. In Section \ref{section_iii}, we provide one of the numerical approaches, which we then apply to the calculation of the vacuum polarization calculation method.

\subsection{Vacuum polarization}


The VP charge density can be derived directly from the fermionic current operator. The expression for the current operator reads \cite{Schwinger1951}

\begin{equation}
    j^\mu (x) = \frac{e}{2} \comm{\Bar{\psi}(x)\gamma^\mu}{\psi(x)}.
    \label{cs_current}
\end{equation}

\noindent This expression originates from the charge symmetrization and is typical for bound-state QED. 
The induced charge density can be found as the timelike component of this current (See \cite{Wichmann1956}), if we 
consider the solutions for wave functions in an external field (\ref{wave_1}-\ref{wave_2}). However, here, we shall
find the same result with a more general approach, i.e. from the scattering amplitude, using the expression for the energy correction
from \cite{Soff1988, Mohr1998}. Later, we shall employ this expression to calculate the Wichmann-Kroll correction. With the help of the left diagram in 
Fig. \ref{fig:vp_diag} and Feynman rules for bound-state QED, we obtain the expression for the energy correction to the $n$-th level electron 
associated with the vacuum polarization:

\begin{widetext}
\begin{equation}
    \Delta E_n = 4\pi i \alpha \int \dd(t_2-t_1) \int \dd^3 x_2 \int \dd^3 x_1 \Bar{\phi}_n(x_2) \gamma^\mu \phi_n(x_2) D_F(x_2-x_1) \Tr[\gamma_\mu S_F (x_1,x_1)],
    \label{engy_amp}
\end{equation}
\end{widetext}

\noindent where

\begin{equation}
    D_F(x_2-x_1) = -i \int \frac{\dd^4 k}{(2\pi)^4} \frac{e^{-ik(x_2-x_1)}}{k^2 + i\varepsilon}
\end{equation}

\noindent is the photon propagator, and the 
fermion propagator is defined as

\begin{equation}
\begin{aligned}
    S_F(x_2,&x_1) = \bra{0} \T \{\psi(x_2)\Bar{\psi}(x_1)\}\ket{0} \\
    &= \begin{cases}
      \sum\limits_{E_n > 0} \phi_n(x_2) \Bar{\phi}_n(x_1), & \text{$t_1<t_2$}\\
      -\sum\limits_{E_n < 0} \phi_n(x_2) \Bar{\phi}_n(x_1), & \text{$t_1>t_2$}
    \end{cases}  \\
    &= \frac{1}{2\pi i} \int\limits_{C_F} \dd z \sum\limits_n \frac{\phi_n(\vb{x}_2) \Bar{\phi}_n(\vb{x}_1)}{E_n - z(1+i\delta)}e^{-z(t_2-t_1)} \\
    &= \frac{1}{2\pi i} \int\limits_{C_F} \dd z \,G(\vb{x}_2,\vb{x}_1,z(1+i\delta)) \gamma^0 e^{-z(t_2-t_1)}.
    \label{ferm_prop_def}
\end{aligned}
\end{equation}

\noindent Above, $G(\vb{x}_2,\vb{x}_1,z(1+i\delta))$ is the 
Green's function of the time-independent Dirac equation (\ref{dirac_eq_2}), and $\phi_n(x)$ are the solutions in the external field. The second line of (\ref{ferm_prop_def}) is used 
in our FBS method, while the last two lines are used for the direct integration method \cite{Soff1988, Mohr1998}.

We evaluate integrals in (\ref{engy_amp}) and, assuming that the system is stationary, find the energy correction to the $j$-th electron energy level:

\begin{equation}
    \Delta E_j = \int \dd^3 x_1 \phi_j^\dagger (\vb{x}_1) U(\vb{x}_1) \phi_j(\vb{x}_1),
    \label{en_cor}
\end{equation}

\begin{equation}
\begin{aligned}
    U(\vb{x}_2) &= \frac{i\alpha}{2\pi} \int \dd^3x_1 \frac{1}{\abs{\vb{x}_1 - \vb{x}_2}} \int\limits_{C_F} \dd z \Tr G(\vb{x}_1, \vb{x}_2, z) \\
    &= \frac{i\alpha}{2\pi} \int \dd^3x_1 \frac{1}{\abs{\vb{x}_1 - \vb{x}_2}} \Tr[\gamma_0 S_F (x_1,x_1)].
    \label{eff_pot}
\end{aligned}
\end{equation}

We compare equation (\ref{eff_pot}) with the Coulomb law 

\begin{equation}
    U(\vb{x}) = -e \int \dd^3 x' \frac{\rho(\vb{x}')}{\abs{\vb{x}-\vb{x}'}},
\end{equation}

\noindent and finally, obtain 
the expression for the induced charge density \cite{Schwinger1951, Wichmann1956}:

\begin{equation}
\begin{aligned}
    &\rho(\vb{x}) = ie \left.\Tr[S_F(x,x')\gamma_0]\right|_{x \rightarrow x'} \\
    &= \frac{ie}{2} \left(\sum\limits_{E_n > 0} \phi_n(\vb{x}) \phi^\dagger_n(\vb{x}) \right. - \left. \sum\limits_{E_n < 0} \phi_n(\vb{x}) \phi^\dagger_n(\vb{x})  \right).
    \label{ind_charge}
\end{aligned}
\end{equation}

\noindent Above, $x \rightarrow x'$ should be understood as an average over the limits from the left and from the right \cite{Schwinger1951, Mohr1998}. On the other hand, if we start from the current (\ref{cs_current}), similar to 
\cite{Schwinger1951}, we 
find:

\begin{equation}
    j^\mu(\vb{x}) = ie \left.\Tr{S_F(x,x')\gamma^\mu}\right|_{x \rightarrow x'},
\end{equation}

\noindent where the timelike component is, once again, the charge density.

\section{Finite basis set approximation}\label{section_iii}

One of the approaches to the variational problem for the radial Dirac equation is the Ritz-Rayleigh or Ritz-Galerkin method. 
The idea is to use the 
finite set of basis functions, and approximate the solution to the Dirac equation as a sum of these functions with weight coefficients found from a variation problem. 
The energy spectrum is found as a byproduct of this procedure.

\subsection{Variational method}

The core of the finite basis method is the approximation of the exact solution to the radial Dirac equation with the weighted sum over a finite set of some basis functions. Unknown coefficients in this sum are found with the help of variational methods \cite{Grant1986, grant2007relativistic, Grant2000, Johnson1988, Sapirstein1996}. 

In this method, the energy expectation value (or action in some papers, alternatively) is expressed using the "test" wave function, defined as the sum over the given finite basis functions with unknown coefficients. Next, these coefficients are found by variation of the energy expectation value \cite{Drake1981} (or the action \cite{Johnson1988}).

Let us define the test function next way:

\begin{equation}
    \Psi(r) = \sum\limits_{i=1}^n v_i \pi_i(r),
    \label{test_fun}
\end{equation}

\noindent where $\pi_i(r)$ are functions from some known finite basis set. The expectation value (or the Rayleigh quotient) for $\varepsilon$, which happens 
to be the 
upper bound for the true value of energy \cite{Drake1981} reads:

\begin{equation}
    \varepsilon = \frac{\bra{\Psi}H\ket{\Psi}}{\braket{\Psi}{\Psi}} = \frac{\sum v_i^* v_j H_{ij}}{\sum v_i^* v_j C_{ij}}.
\end{equation}

\noindent We vary the coefficients and require 
the variation of the energy expectation value to be 
minimal. Varying $\varepsilon$, we obtain

\begin{align}
    \pdv{\varepsilon}{v_k^*} =& \frac{\sum v_j H_{kj}}{\sum v_i^* v_j C_{ij}} - \frac{\sum v_i^* v_j H_{ij}\sum v_j C_{kj}}{\left(\sum v_i^* v_j C_{ij}\right)^2} \\
    =& \frac{\sum v_j \left(H_{kj} - \varepsilon C_{kj}\right)}{\sum v_i^* v_j C_{ij}}.
    \label{var_eq}
\end{align}

\noindent  To obtain the 
minimum, these derivatives (\ref{var_eq})  should be 
equal to zero, which leads to secular equations:

\begin{equation}
    v_j \left(H_{kj} - \varepsilon C_{kj}\right) = 0,
\end{equation}

\noindent or, in matrix form:

\begin{equation}
    \vb{H} \vb{v} = \varepsilon \vb{C} \vb{v}
    \label{eig_eq}.
\end{equation}

Above, we recognize a generalized eigenvalue problem with hermitian matrices $\vb{H}$ and $\vb{C}$, where $\varepsilon$ are the upper bounds for the Dirac equation spectrum, and $v$ are eigenvectors, consisting of coefficients involved in 
the sum that defines the wave functions (\ref{test_fun}). With 
the increase in the size of the basis set, 
the approximate spectrum values $\varepsilon$ (and the wave functions calculated using the obtained 
coefficients) become closer to their true values. \footnote{There is an important feature of eigenvectors, sometimes called $\vb{B}$-orthogonality. This means 
that if in a $\vb{A}\vb{v}=\lambda\vb{B}\vb{v}$ problem, the matrices $\vb{A}$ and $\vb{B}$ are both symmetric or Hermitian, and if $\vb{B}$ is a positive-definite matrix, then $\vb{v}_i^T \vb{B} \vb{v}_j = \delta_{ij}$, or, in our notation, $\vb{v}_i^T \vb{C} \vb{v}_j = \delta_{ij}. $ Here we must note 
that since eigenvectors are defined up to a 
constant factor, they can vary for different calculation methods and programs; as a result, 
diagonal elements on the 
right side in the equations above can differ from one. Methods like \texttt{DSYGV} return such eigenvectors, which give $\delta_{ij}$ in the $\vb{B}$-orthogonality relations. This property allows control 
over the calculation precision, since in numerical calculation, one would see $\vb{v}_i^T \vb{C} \vb{v}_i = 1 \pm \varepsilon$, where $\varepsilon$ is a 
small number, which become larger if the solution error increases.  }


In the case of the radial Dirac equation (\ref{dirac_eq_mat}), we solve for the ``large'' and ``small'' 
components of the wave function, therefore, we define two sums: 

\begin{align}
    P_\kappa &= \sum\limits_{i=1}^n p_i \pi^+_i(r) \\
    Q_\kappa &= \sum\limits_{i=1}^n q_i \pi^-_i(r),
\end{align}

\noindent where $\pi^\pm_i$ are two basis sets, which could be identical 
or different, depending on the researcher's choice. With this definition, 
$\vb{H}$ can be written in the following form: 

\begin{equation}
    \vb{H} = \begin{pmatrix}
        \vb{H}^{LL} & \vb{H}^{LS} \\
        \vb{H}^{SL} & \vb{H}^{SS}
    \end{pmatrix}   
\end{equation}


\noindent where $L,S$ stands for ``large'' and ``small'' 
components. For instance, 
for the first quadrant, we have

\begin{equation}
    H^{LL}_{ij} = \int\limits_0^\infty \pi^+_i(r)\pi^+_j(r) \left(m + V(r)\right) \dd r,
\end{equation}

\noindent and analogously for the other quadrants. For the correct solution, we need to take into account boundary 
conditions \cite{Johnson1988, Sapirstein1996}. This can be done by adding a special term to the matrix $\vb{H}$ \cite{Johnson1988}, or by choosing basis functions that 
satisfy the boundary conditions 
for any set of coefficients (for example, 
equal to zero at $r=0$ and $\infty$).

It is convenient to unite coefficients in
$P_\kappa, Q_\kappa$  like
$v_i = (p_1,p_2, ... p_n, q_1, q_2, ... , q_n)$ \cite{Johnson1988}:

\begin{align}
    P_\kappa &= \sum\limits_{i=1}^n v_i \pi^+_i(r) \label{basis_fncs1}\\
    Q_\kappa &= \sum\limits_{i=n+1}^{2n} v_i \pi^-_{i-n}(r)
    \label{basis_fncs2}
\end{align}


\subsection{Dual-kinetic balance}

There is a well-known problem with the na\"ive choice of basis functions $\pi^\pm$ (when the large and small components are independent) in the Dirac problem. Spurious nonphysical states emerge in the spectrum \cite{Tupitsyn2008}. Such states can be ignored \cite{Sapirstein1996, Drake1981}. However, it is more convenient to choose a basis set that mitigates these spurious states.

In the so-called kinetic balance (KB) approach, 
large component solutions are tied 
to the small components solutions: $\pi^+=\frac{1}{2}(\frac{\dd}{\dd r} + \frac{\kappa}{r})\pi^-$. However, this approach is asymmetrical for positive and negative energy terms, which can, for example, lead to a violation of $\mathcal{C}$-symmetry in vacuum polarization calculations \cite{Salman2023}. In the 
alternative 
dual-kinetic balance (DKB) approach \cite{Shabaev_DKB}, 
positive and negative energy components are mixed symmetrically, and the 
approximate solution is written as 
(compare with (\ref{basis_fncs1}, \ref{basis_fncs2})):

\begin{equation}
\begin{aligned}
    \varphi_\kappa = \begin{bmatrix}
        P_\kappa  \\
        Q_\kappa 
  \end{bmatrix}
  & = \sum\limits_{i=1}^n
    v_i\begin{bmatrix}
    \pi^{+}_i  \\
    \frac{1}{2}\left(\frac{d}{dr} + \frac{\kappa}{r}\right)\pi^{+}_i 
  \end{bmatrix} \\&+
  \sum\limits_{i=n+1}^{2n}
    v_i\begin{bmatrix}
    \frac{1}{2}\left(\frac{d}{dr} - \frac{\kappa}{r}\right)\pi^{-}_{i-n}  \\
    \pi^{-}_{i-n} 
  \end{bmatrix}.
\end{aligned}
\end{equation}


With DKB, 
for example, 
the overlap matrix in (\ref{eig_eq}) becomes

\begin{equation}
    \vb{C} = \begin{pmatrix}
        \vb{C}^{LL} & \vb{C}^{LS} \\
        \vb{C}^{SL} & \vb{C}^{SS}
    \end{pmatrix},
\end{equation}

\begin{align}
    C^{LL}_{ij} =& \int_0^\infty \pi^{+}_i \pi^{+}_j dr \\
    C^{LS}_{ij} =& \frac{1}{2}\int_0^\infty \pi^{+}_i \left\{\left(\frac{d}{dr} - \frac{\kappa}{r}\right)\pi^{-}_j \right\} dr \\
    C^{SS}_{ij} =& \frac{1}{4}\int_0^\infty \left\{\left(\frac{d}{dr} - \frac{\kappa}{r}\right)\pi^{-}_i \right\} \left\{\left(\frac{d}{dr} - \frac{\kappa}{r}\right)\pi^{-}_j \right\} dr \\
    \vb{C}^{SL} =& (\vb{C}^{LS})^T.
\end{align}

\noindent The expressions for $\vb{H}$ matrix can be found similarly.

\subsection{Induced charge density in the finite basis set approximation}

Now we turn to the definition of the vacuum polarization 
charge density, using an FBS approach. In \cite{Salman2023}, it was suggested to calculate the induced charge density with the finite basis set approximation as follows: 
the charge density (\ref{ind_charge}) is obtained as a 
sum of wave functions up to number $2n$, which are found by solving the Dirac equation.


In \cite{Wichmann1956}, it was shown that the vacuum polarization charge density can be decomposed into components with different angular momentum states:

\begin{equation}
    \rho(\vb{x}) = \sum\limits_{\kappa = \pm 1}^{\pm\infty} \rho_\kappa(\vb{x}) = \frac{e}{2\pi i} \int\limits_{C_F} \dd z  \sum\limits_{\kappa \pm 1}^\infty \frac{\abs{\kappa}}{2\pi} \Tr G_\kappa(\vb{x}_1, \vb{x}_2, z).
    \label{vp_kappa_dec}
\end{equation}

\noindent If the spectrum is assumed to be discrete, each of the $\rho_\kappa$ components is 
defined as \cite{Salman2023}

\begin{align}
    \rho_\kappa(\vb{x}) =& \frac{\abs{\kappa}}{2\pi}\frac{e}{2}\frac{1}  {r^2}\sum\limits_n \text{sgn}(E_{\kappa,n})\rho_{\kappa,n}(r),
    \label{rho1}
    \\ \rho_{\kappa,n}(r) &= \varphi_{\kappa,n}^\dagger\varphi_{\kappa,n} = P_{n,\kappa}^2 + Q_{n,\kappa}^2.
\end{align}


For the exact solution, the sum in (\ref{rho1}) should be infinite, but an approximate solution can be found with a finite basis set. In the latter case, the summation goes from 
$i=1$ 
to $2n$, where $n$ is the 
number of finite basis set functions.

The VP charge density contains divergent terms such as the 
Uehling term, which are often treated separately, since the renormalized expression for the Uehling potential is known \cite{Mohr1998}. To exclude 
this divergence, we expand the charge density in the 
powers of $Z$:

\begin{equation}
    \rho_\kappa(r,Z) \equiv \sum\limits_{k=0}^\infty \rho_\kappa^{(k)}(r,Z) = \sum\limits_{k=0}^\infty \left.\pdv[k]{}{Z}\rho_\kappa(r,Z)\right|_{Z=0} \frac{Z^k}{k!}.
\end{equation}

\noindent Here, $\rho_\kappa^{(1)}$ is the Uehling term, which contains linear divergence. Following \cite{Soff1988, Mohr1998}, we separate the finite part $\rho^{n \geq 3}$ by subtracting the diverging terms from the total induced charge density. Apart from the Uehling term, there is a spurious diverging piece in the third-
order term $\Tilde{\rho}^{(3)}$, which 
should also be considered: 

\begin{equation}
    \rho^{n \geq 3} = \rho - \rho^{(1)} - \Tilde{\rho}^{(3)}
\end{equation}

\noindent However, since we sum over finite number of angular momentum terms, the $\Tilde{\rho}^{(3)}$ part vanishes, as shown in, for example, \cite{Soff1988}. Then, subtracting the linear term, we have \cite{Rinker1975, Salman2023}

\begin{equation}
    \rho_\kappa^{n \geq 3}(r,Z) = \rho_\kappa(r,Z) - \lim\limits_{\delta \rightarrow 0} \frac{Z}{\delta} \rho_\kappa(r,\delta)
\end{equation}

Moreover, as discussed 
in \cite{Salman2023}, 
$\mathcal{C}$-symmetry is important 
to obtain the correct numerical solution for VP charge density. 
If this symmetry is not ensured in the finite basis set calculation method, the results can be corrupted. 
To provide the $\mathcal{C}$-symmetry is obeyed, it can be enforced directly: 

\begin{equation}
    \rho_{\kappa,\mathcal{C}}(r,Z) = \frac{1}{2}\left(\rho_\kappa(r,Z) - \rho_\kappa(r,-Z)\right),
\end{equation}

\noindent so that

\begin{equation}
\begin{aligned}
    \rho_{\kappa,\mathcal{C}}^{n \geq 3}(r,Z) &= \frac{1}{2}\left(\rho_\kappa^{n \geq 3}(r,Z) - \rho_\kappa^{n \geq 3}(r,-Z)\right) \\
    &- \frac{Z}{2\delta}\left(\rho_\kappa^{n \geq 3}(r,\delta) - \rho_\kappa^{n \geq 3}(r,-\delta)\right).
    \label{c_sym}
\end{aligned}    
\end{equation}

It was discussed in
\cite{Salman2023}, that 
the KB basis generally violates the $\mathcal{C}$-symmetry, while the DKB basis does not. We shall
use (\ref{c_sym}) in our calculations below.

\subsection{Basis sets}

In the finite basis approximation method, there are several 
basis sets commonly 
used for bound-state QED calculations. One of these 
is the B-spline basis, 
widely used for the self-energy diagrams \cite{Shabaev_DKB}. However, this basis does not work well for the vacuum polarization calculations. We show it below.

The B-spline basis is defined as: 
\cite{Johnson1988, Sapirstein1996, deboor1978}

\begin{equation}
    B_{i,1}(x) = \begin{cases}
      1, & \text{$t_i\leq x < t_{i+1}$}\\
      0, & \text{otherwise}
    \end{cases}
\end{equation}

\noindent where $\{t_1, ... , t_n\}$ are the knot points, and higher degree B-splines are defined by the recurrent formula:

\begin{equation}
\begin{aligned}
    B_{i,k}(x) =& \frac{x - t_i}{t_{i+k} - t_{i+1}}B_{i,k-1}(x) \\
    &+ \frac{t_{i+k} - x}{t_{i+k} - t_{i+1}} B_{i+1,k-1}(x).
\end{aligned}
\end{equation}

\noindent These 
functions are $k$-degree polynomials 
defined on a finite interval 
and equal to zero elsewhere. B-spline basis functions are well-suited for integrating using Gauss-Legendre quadrature, and 
linear dependency is not a serious problem in this basis.

Another basis set, 
widely used in quantum chemistry and atomic physics,
is the Gaussian basis set with the following basis functions 
\cite{Grant1986, Grant1988, Salman2023}:

\begin{equation}
    \pi^{\pm}_i(r) = \mathcal{N} r^{d_\pm} e^{-\zeta_i r^2},
    \label{gauss_basis}
\end{equation}

\noindent where $d_\pm$ are chosen to simulate the wave function behavior near $r=0$ and defined depending on the type of the potential, 
and $\mathcal{N}$ is the normalization factor. For point-like (Coulomb) potential, $d_\pm$ would be equal to $\gamma=\sqrt{\kappa^2 - (Z\alpha)^2}$ \cite{berestetskii}, but for shell model of nucleus, the potential near $r=0$ is constant, which corresponds to 
free-particle behavior ($Z=0$), thus 
in this case, we have \cite{grant2007relativistic}

\begin{equation}
    d_\pm = \abs{\kappa \pm \frac{1}{2}} + \frac{1}{2}.
\end{equation}

The coefficients $\zeta_i$ in (\ref{gauss_basis}) are generated to form a geometric sequence \cite{Grant1988}:

\begin{equation}
    \zeta_i = \zeta_1 (\zeta_n/\zeta_1)^{\frac{i-1}{n-1}},
\end{equation}


\noindent where $\zeta_{1,n}$ are 
the first and last coefficients, and $n$ is the size of the basis set. The Gaussian basis is better suited for 
vacuum polarization calculations, as we shall
show below. However, this basis is subject to the 
linear dependence problem, 
which makes increasing the basis set size problematic. We shall also discuss this issue below.

\section{Results}\label{section_iv}

In this section, we calculate the vacuum polarization charge density and the Wichmann-Kroll correction 
associated with this density for 
hydrogen-like levels of 
various heavy ions. We plot the VP-induced charge density and compare it 
to the results of \cite{Soff1988, Mohr1998}. We also compare the obtained Wichmann-Kroll corrections 
with the corresponding results from \cite{Persson1993}.

\subsection{Calculation of VP charge density}\label{sec_calc_vp_density}

First, we calculated VP charge densities 
as a function of the distance from the nucleus $r$. To calculate this density, we need to sum over an infinite number of angular momentum components. However, for heavy elements, a 
sum over partial components $\kappa$ converges fast enough, and 
the first few terms can be sufficient. Following \cite{Mohr1998}, we denote

\begin{equation}
    \rho_{\abs{\kappa}} = \rho_{-\kappa} + \rho_{\kappa}.
\end{equation}

\noindent In VP calculations, partial terms are taken in pairs $\rho_{-\kappa} + \rho_{\kappa}$ \cite{Wichmann1956, Mohr1998}, which is necessary for cancellation of 
the zero-potential contribution \cite{Persson1993, Salman2023}.

\begin{figure}[!htb]
    \centering
    \includegraphics[width=1\linewidth]{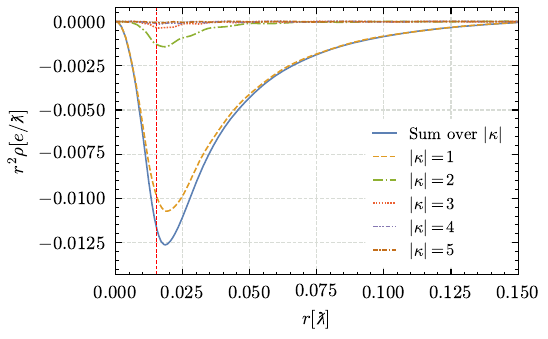}
    \caption{Radial vacuum polarization charge density of order $\alpha(Z\alpha)^{n \geq 3}$ for $Z=92$ with nuclear radius $r_n = 5.751$ fm, at a small 
    distance from the nucleus. Calculations were performed in Gaussian basis with the basis set size $n=30$. Dashed lines: contributions from different $\kappa$; solid line: sum over the first five 
    contributions from different $\kappa$. A vertical line denotes the nuclear radius.
    }
    \label{fig:vp_plot}
\end{figure}

\begin{figure}[!htb]
    \centering
    \includegraphics[width=1\linewidth]{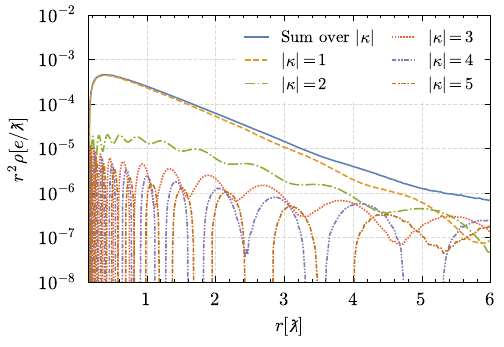}
    \caption{Same as Fig. \ref{fig:vp_plot}, but in log-scale, at a large distance from the nucleus.
    }
    \label{fig:vp_plot_log}
\end{figure}

\begin{figure}[!htb]
    \centering
    \includegraphics[width=1\linewidth]{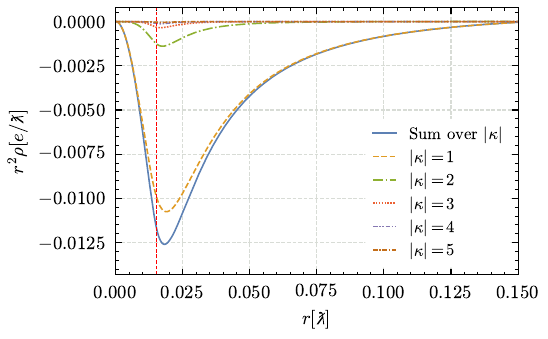}
    \caption{Same as Fig. \ref{fig:vp_plot}, but with the basis set size $n=100$.
    }
    \label{fig:vp_plot_mp}
\end{figure}

\begin{figure}[!htb]
    \centering
    \includegraphics[width=1\linewidth]{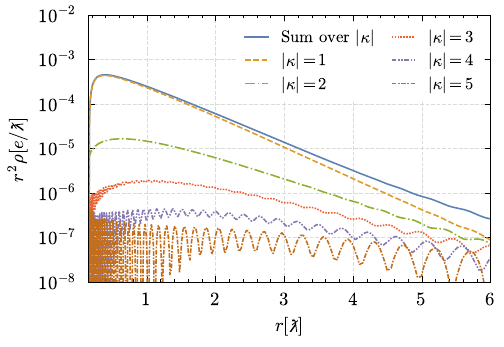}
    \caption{Same as Fig. \ref{fig:vp_plot_log}, but with the basis set size $n=100$.
    }
    \label{fig:vp_plot_log_mp}
\end{figure}

\begin{figure}[!htb]
    \centering
    \includegraphics[width=1\linewidth]{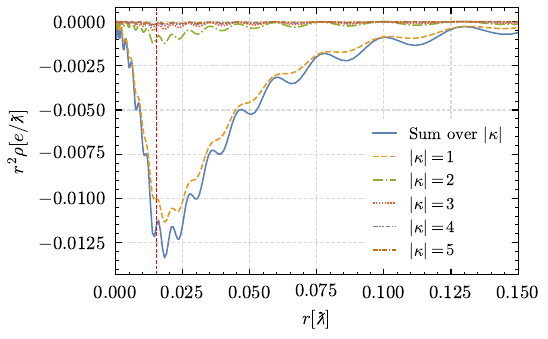}
    \caption{Radial vacuum polarization density of order $\alpha(Z\alpha)^{n \geq 3}$ for $Z=92$ with nuclear radius $r_n = 5.751$ fm. Calculations were performed in the B-spline basis. Dashed lines: contributions from different $\kappa$; solid line: sum over the first five 
    contributions from different $\kappa$.
    }
    \label{fig:vp_bs_plot}
\end{figure}

\begin{figure}[!htb]
    \centering
    \includegraphics[width=1\linewidth]{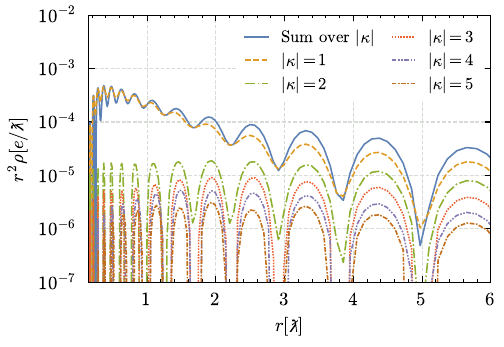}
    \caption{Same as Fig. \ref{fig:vp_bs_plot}, but in log-scale.
    }
    \label{fig:vp_bs_plot_log}
\end{figure}

In 
Figs.~\ref{fig:vp_plot}-\ref{fig:vp_bs_plot_log}, we show VP densities calculated for hydrogen-like Uranium $Z=92$ with the nuclear radius 
$r_n = 5.751$ fm used different basis sets. We use a shell-like nucleus model 
throughout this work, where the nucleus charge density is modeled by

\begin{equation}
    \rho_n (r) = \delta (r - r_n). 
    \label{shell}
\end{equation}

\noindent This nucleus model was used in \cite{Mohr1998, Salman2023}; it is 
simple and 
provides a better behavior of the induced charge density near $r=0$ than a point-like charge with $ \rho_n (\vb{x}) = \delta (\vb{x})$.

We performed our calculations using both the B-spline and Gaussian basis sets. For the Gaussian basis, the matrices for the eigenvalue problem equation (\ref{eig_eq}) can be calculated analytically for point- and shell-like potentials, which makes calculations more efficient and precise. We calculated these matrices using both analytical and numerical integration. For the B-spline basis set, 
these matrices were calculated 
using the Gauss-Legendre quadrature. We used parameters $\zeta_1 = (\frac{1}{386.159})^2 \cross 10^{3}$ and $\zeta_n = (\frac{1}{386.159})^2 \cross 10^{11}$, and the sizes of the Gaussian basis sets were 
chosen equal to 30 and 100. For the B-spline set, we used a basis set with 40 
basis functions
, and for knot 
values, we used a power 
grid with denser knots 
inside the nucleus.

There is a well-known problem associated with the Gaussian basis set (as opposed 
to the B-spline basis): 
as its size increases, the basis functions can become efficiently linearly dependent. This imposes an 
upper limit on 
the size of a Gaussian basis set; upon 
reaching this limit, 
the numerical solution collapses. This limit depends mainly on the numerical precision; thus, 
for our parameters, we can use only 
30 basis functions if \texttt{float} numbers are used. This limit can be increased by using 
arbitrary precision numbers (at the expense 
of performance speed). We used 70-digit numbers for calculations with a set of 
100 basis functions.


For our calculations, we used a Python program with \texttt{mpmath} package for multiple-precision calculations \cite{mpmath}. In Table \ref{tab:benchmark}, we present the approximate evaluation time 
required for the calculation of $\rho_\abs{\kappa}(r)$ for one value of $\abs{\kappa}$. The algorithm is quite time-efficient 
when machine-precision (\texttt{float}) numbers are used, while for arbitrary precision calculation, the required time is still tolerable (but a lot larger than for float, since symbolic calculations are used). \footnote{We note that the evaluation
time can be reduced 
by choosing a smaller grid, less digit numbers or by overall optimizing the calculation algorithm, which was not our main goal in current work.}

\begin{table}[!htb]
\footnotesize
\caption{\label{tab:benchmark}%
Approximate time for the 
calculation of one $\rho_\abs{\kappa}(r)$ term, for different number formats and sizes of the Gaussian basis set.
}
\begin{ruledtabular}
\begin{tabular}{c|c}
Basis set size, number format & Evaluation time, approximate   \\
\hline n=30, float & 2 seconds  \\
n=100, 70-digit multiple precision & 10 minutes 
\\
\end{tabular}
\end{ruledtabular}
\end{table}

Figures~\ref{fig:vp_plot}, \ref{fig:vp_plot_log} show the VP charge density calculated using a Gaussian basis set with $n=30$. Here, the DKB method was applied. The radius scale
is given in Compton wavelengths units $\lambdabar = \frac{1}{m}$. The presented plots are 
in good agreement with the results of 
\cite{Soff1988, Mohr1998, Salman2023} for $\abs{\kappa}=1$, which gives the major part of the total $\rho^{n\geq 3}$ and $\Delta E_\text{WK}$ value. However, there are 
noticeable oscillations, which spoil the charge density curves for 
higher values of $\abs{\kappa}$ or, equivalently, for $r^2\rho^{n \geq 3}$ 
smaller than $\approx 10^{-5}$. In Figs.~\ref{fig:vp_plot_mp}, \ref{fig:vp_plot_log_mp}, we show 
calculation results for the same problem, but with the basis set size $n=100$. Here, we do not use the DKB method, since imposing 
$\mathcal{C}$-symmetry by the rule (\ref{c_sym}) seems to be sufficient. Evidently, 
the results are much better this time, especially at a 
large distance from the nucleus. In Figs.~\ref{fig:vp_bs_plot}, \ref{fig:vp_bs_plot_log}, we show VP plots 
calculated using the B-spline basis set. This time, the oscillations are 
significant for all curves corresponding to different orders in $\kappa$.

\subsection{Wichmann-Kroll correction}

We have obtained the 
VP charge density, and now 
we can calculate the corresponding correction to the atomic energy spectrum. 
We calculate 
this correction 
to test the accuracy of our VP charge density calculations and analyze the applicability of the proposed method to the Wichmann-Kroll correction calculations. 
As we have shown in Section \ref{sec_calc_vp_density}, 
a Gaussian basis set used for VP calculations provides 
better results, and therefore, we use these basis functions to calculate the Wichmann-Kroll energy corrections. The major part of the energy shift 
caused by the vacuum polarization 
comes from the $Z\alpha$- order correction associated with the Uehling potential (see, for example, \cite{Persson1993}). The correction originating from the higher orders is called the Wichmann-Kroll correction \cite{Wichmann1956}, and it can be calculated using $\rho^{n \geq 3}$. The potential 
created by this charge density 
is \cite{Mohr1998}:


\begin{equation}
    V(r) = -4\pi\alpha \left[\frac{1}{r}\int\limits_0^r \rho(r') r'^2 \dd r' +  \int\limits_r^\infty \rho(r') r' \dd r'\right],
\end{equation}

\noindent from which, using (\ref{en_cor}), we can 
find the energy correction for the electron with quantum numbers $n$ and $\kappa$:


\begin{equation}
\begin{aligned}
    \Delta E_{\kappa, n} &= \int \varphi_{\kappa,n}^\dagger(r) V(r) \varphi_{\kappa,n}(r) \dd r =  \\
    & \int V(r) \left(P_{n,\kappa}^2 + Q_{n,\kappa}^2\right) \dd r = \bra{n,\kappa}V\ket{n,\kappa}
\end{aligned}
\end{equation}

Since the VP charge density can be expanded into different angular momentum contributions (\ref{vp_kappa_dec}), these terms contribute 
to individual energy corrections terms, which are summed 
to obtain the total value:

\begin{equation}
    \Delta E_{n} = \sum\limits_\kappa \Delta E_{\kappa, n}
    \label{energy_conv}
\end{equation}

\noindent Because this value converges quite fast with increasing 
$\abs{\kappa}$, we can take 
only the first few values of $\Delta E_{\kappa, n}$ to get accurate results (an extrapolation can be used \cite{Persson1993}, but the precision of our method is not 
high enough for us to consider this improvement). In this paper, we consider the first 5 terms in $\abs{\kappa}$.

To compare the FBS approach for the calculation of 
the VP charge density with the traditional ones, we also performed calculations using a standard approach based on 
integration of Green's functions (see, 
for example, \cite{Mohr1998}). In Table \ref{tab:coef_conv}, we show the results for Wichmann-Kroll correction 
components for different $\abs{\kappa}$, calculated with standard Green's function integration and using the FBS approximation with Gaussian basis sets of sizes 30 and 100. The calculations were performed for hydrogen-like Uranium with $Z=92$, $r_n=5.751$ fm, using the shell nucleus model 
(\ref{shell}). For higher $\abs{\kappa}$, the corresponding terms should converge, but for the FBS calculation, the oscillations become considerable; therefore, higher $\kappa$ terms vanish slower than they should, which adds to the error of the FBS method. Moreover, from Table \ref{tab:coef_conv}, we can see 
that the correction value converges to the value 
calculated by the integration method 
with increasing size of the basis set. 

\begin{table}[!htb]
\footnotesize
\caption{\label{tab:coef_conv}%
Contributions from different $\kappa$ 
to the Wichmann-Kroll correction and their sum for 1s electron in hydrogen-like Uranium for different calculation methods. In parentheses, the difference between the corrections calculated by FBS and Green's function integration methods is shown. 
}
\begin{ruledtabular}
\begin{tabular}{c|c|c|c}
 &  \multicolumn{3}{c}{$\Delta E_{\text{WK}}$ ($\Delta E^{\text{FBS}}_{\text{WK}}$ - $\Delta E^{\text{Green}}_{\text{WK}})$, eV} \\
\hline $\abs{\kappa}$ & Green's integr. & FBS $n=30$ & FBS $n=100$ \\
\hline 1 & 4.473
 & 4.479 (0.006) & 4.489 (0.016) \\
2 & 0.394 & 0.467 (0.073) & 0.405 (0.011)
\\ 3 & 0.081 & 0.145 (0.064) & 0.089 (0.008)
\\ 4 & 0.024 & 0.059 (0.035) & 0.032 (0.008)
\\ 5 & 0.009 & 0.046 (0.037) & 0.018 (0.009)
\\
Sum & 4.99 & 5.198 (0.208) & 5.033 (0.043)

\\
\end{tabular}
\end{ruledtabular}
\end{table}

Further, we calculated Wichmann-Kroll corrections for the first few electron orbitals of several elements. 
The results are presented in Tables \ref{tab:table1} and \ref{tab:table2}. We compare our results with \cite{Persson1993}. As can be seen from these tables, 
the FBS calculations 
provide
results with a considerable error compared to the 
results from \cite{Persson1993}. However, for the first two orbitals ($1s_{1/2}$, $2s_{1/2}$), our method gives results with error $\lesssim$ 10\% for the basis set size $n=30$ and $\lesssim$ 3\% for the basis set size $n=100$. Again, similar to 
Table \ref{tab:coef_conv}, the error decreases with 
increasing size of the basis set. 
In Table \ref{tab:table2}, we show only calculations with $n=100$, since for the basis set with $n=30$, the error here is too high.


\begin{table*}[!htb]
\caption{\label{tab:table1}%
Wichmann-Kroll vacuum polarization effects, calculated using FBS method for various elements, in eV, and error $(\Delta E_{\text{FBS}} - \Delta E_{\text{Persson}})/\Delta E_{\text{Persson}}$, \%. For comparison, results from Persson et al. (1993) \cite{Persson1993} are given. Calculations were made with basis set sizes $n=30$ and $n=100$. The nuclear radii of the elements 
are 5.273, 5.505, 5.860, and 5.976 fm, respectively. Here, 
$1s_{1/2}$ and $2s_{1/2}$ orbitals are considered.
}
\begin{ruledtabular}
\begin{tabular}{l|c|c|c|c|c|c}
 &\multicolumn{3}{c}{$1s_{1/2}$} & \multicolumn{3}{c}{$2s_{1/2}$}\\
\cline{1-7}
 & FBS n=30 & FBS n=100 & Persson(1993) & FBS n=30 & FBS n=100 & Persson(1993) \\
\hline
$\prescript{}{70}{\text{Yb}}$ & 0.8462 (2.16\%)
 & 0.8375  (1.10\%)
 & 0.8283  & 0.1339 (11.77\%)

 & 0.1236 (3.17\%) & 0.1198 \\
$\prescript{}{82}{\text{Pb}}$ & 2.3575 (2.95\%)
 & 2.3087 (0.82\%) & 2.2900 & 0.3850 (8.94\%) & 0.3576 (1.19\%) & 0.3534
 \\
$\prescript{}{92}{\text{U}}$ & 5.1982 (4.25\%) & 5.0332 (0.94\%) & 4.9863 & 0.8381 (2.03\%)
 & 0.8293 (0.96\%) & 0.8214 \\
$\prescript{}{100}{\text{Fm}}$ & 9.378 (3.41\%)
 & 9.1108 (0.46\%) & 9.069 & 1.684 (6.11\%)
 & 1.5964 (0.59\%) & 1.587 \\
\end{tabular}
\end{ruledtabular}
\end{table*}

\begin{table*}[!htb]
\caption{\label{tab:table2}%
Wichmann-Kroll corrections, calculated using FBS method for various elements, in eV, and error $(\Delta E_{\text{FBS}} - \Delta E_{\text{Persson}})/\Delta E_{\text{Persson}}$, \%. For comparison, results from Persson et al. (1993) \cite{Persson1993} are given. Calculations made with basis set size $n=100$. The nuclear radii 
are the same as in Table \ref{tab:table1}. Here, $2p_{1/2}$ and $2p_{3/2}$ orbitals are considered.
}
\begin{ruledtabular}
\begin{tabular}{l|c|c|c|c}
 &\multicolumn{2}{c}{$2p_{1/2}$} & \multicolumn{2}{c}{$2p_{3/2}$}\\
\cline{1-5}
 & FBS n=100 & Persson(1993) & FBS n=100 & Persson(1993) \\
\hline
$\prescript{}{70}{\text{Yb}}$ 
 & 0.0170 (11.11\%)
 & 0.0153 

 & 0.0041 (46.43\%) & 0.0028 \\
$\prescript{}{82}{\text{Pb}}$ 
 & 0.0698 (5.76\%) & 0.0660 & 0.0120 (27.66\%) & 0.0094
 \\
$\prescript{}{92}{\text{U}}$ & 0.2138 (3.94\%) & 0.2057
 & 0.0269 (19.03\%) & 0.0226 \\
$\prescript{}{100}{\text{Fm}}$
 & 0.5091 (2.23\%) & 0.498
 & 0.0497 (15.58\%) & 0.043 \\
\end{tabular}
\end{ruledtabular}
\end{table*}

\section{Discussion and conclusion}\label{section_v}

In this study, we have calculated vacuum polarization (VP) charge densities for several heavy hydrogen-like ions using the finite basis set approximation. Furthermore, we have calculated the Wichmann-Kroll corrections for the first few electron orbitals using these densities. As we show, the calculated charge densities are in good agreement with the results of Soff and Mohr \cite{Soff1988, Mohr1998} for $\abs{\kappa} = 1$. However, this method is subject to oscillation issues (see Figs.~\ref{fig:vp_plot}-\ref{fig:vp_bs_plot_log}), which spoil the results for higher $\abs{\kappa}$ and impose limitations on this calculation method. Conversely, the results become more reliable with larger basis sets. This method can be used to obtain good estimates for VP calculations, or as an independent approximate test for calculations performed by other methods, such as Green's function integration, and can be further improved by optimizing the computation algorithm to use larger basis sets more efficiently or by finding more suitable basis sets.


We have calculated the VP 
charge density using the Gaussian 
and B-spline basis sets, showing that for this task, 
the Gaussian basis provides good results (confirming the calculations from \cite{Salman2023} for $\abs{\kappa}=1$), while B-spline calculations suffer 
from strong oscillations. 

On the other hand, we certainly do not 
state that B-splines cannot be used for vacuum polarization calculations: we only show 
that this basis performs worse than the Gaussian one if the same approach is used. In principle, by using arbitrary precision or 
mixed basis sets, B-splines could be applied 
in VP calculation problems, but this matter should be investigated separately.

A known problem with Gaussian basis sets is linear dependence, which imposes a limit on the maximum size of the basis set. This limitation can be lifted using arbitrary (multiple) precision computation methods. The presented results for the Wichmann-Kroll corrections show the convergence of the finite basis set approximation method with increasing size of the basis set. Such calculations are much slower than standard  \texttt{float}-number calculations, but they are necessary for accurate results when using a large basis set. While calculations with  \texttt{float}-numbers are imprecise (see Tables \ref{tab:table1} and \ref{tab:table2}), they are extremely fast. Our code provides the result in about 2 seconds. This makes the proposed method a good benchmark for testing solutions obtained with time-consuming conventional approaches.
 
We conclude 
that the finite basis approximation method can be successfully used to calculate the Wichmann-Kroll energy corrections, if one is content with an accuracy within a few percent ($\lesssim 3$ \% for a Gaussian basis set with a size of $n=100$ for $1s_{1/2}$ and $2s_{1/2}$ electrons in hydrogen-like heavy ions). We also believe 
that 
the accuracy of this method can be improved further, and it 
could be used 
in various atomic problems, such as evaluation of the higher orders of the Wichmann-Kroll correction or screening effects in lithium-like ions. 


\begin{acknowledgments}
We are grateful to I. Terekhov for criticism. We also deeply appreciate the discussion with M. Salman, who helped to clarify some important details in their paper and gave us useful computation 
tips. The authors thank L. Pogorelskaya for her proofreading of the English manuscript.

The studies in Sec. \ref{section_ii} are supported by the Foundation for the Advancement of Theoretical Physics and Mathematics “BASIS”. The studies in Sec. \ref{section_iii} are supported by the Ministry of Science and Higher Education of the Russian Federation (Agreement No. 075-15-2021-1349). The studies in Sec. \ref{section_iv} are supported by the Russian Science Foundation (Grant No. 22-12-00258; \footnote{\relax https://rscf.ru/en/project/22-12-00258/}).

\end{acknowledgments}

\bigbreak

\bibliography{main.bib}

\end{document}